\documentclass[conference]{IEEEtran}

\ifCLASSINFOpdf
\else
\fi
\usepackage{amsmath}
\usepackage{amsfonts}
\usepackage{amssymb}
\usepackage{graphicx}%
\providecommand{\U}[1]{\protect\rule{.1in}{.1in}}

\begin{document}
%
\title{Cyberbullying and Traditional Bullying in Greece: An Empirical Study}

\author{\IEEEauthorblockN{Maria Papatsimouli,\\ John Skordas, \\Lazaros Lazaridis,\\Eleni Michailidi}
\IEEEauthorblockA{Lab. of Web Technologies \\ 
\& Applied Control Systems \\ Dept. Of Electrical Engineering\\
Western Macedonia Univ. \\ of Applied Sciences,\\ Kozani, Hellas}
\and
\IEEEauthorblockN{Vaggelis Saprikis}
\IEEEauthorblockA{Dept. Of Business Administriation\\
Western Macedonia Univ. \\ of Applied Sciences,\\ Kozani, Hellas}

\and
\IEEEauthorblockN{George F. Fragulis}
\IEEEauthorblockA{Lab. of Web Technologies \\ 
\& Applied Control Systems \\ Dept. Of Electrical Engineering\\
Western Macedonia Univ. of Applied Sciences,\\ Kozani, Hellas}}


%


\maketitle

\begin{abstract}
The rapid evolution of technology is a preferred method of
interacting. A new world was created for young people who are sending emails,
visiting websites, using webcams and chat rooms, and instant messaging through social
media. In the past years, people used face-to-face communication though, in
recent years, people are using internet technology in order to communicate with each other. As a consequence the communication change, created a new type of bullying, cyberbullying, in which bullying is taking place by using internet technology. Cyberbullying is a phenomenon which is increasing day by day over the world. This was one of the reasons that prompted us to do this survey.  Furthermore, bullies’ aggressive reaction is affected by four factors which are conductive during childhood.

\end{abstract}

\begin{IEEEkeywords}
Cyberbullying, traditional bullying, victims, adolescents, victimization, Bullies.

\end{IEEEkeywords}

%
\IEEEpeerreviewmaketitle

\section{Introduction}

According to \cite{Whitney}, and \cite{Olweus1997}  Bullying is described as an aggressive or intentional act or behavior that is carried out
in a repeatedly way against a victim who cannot defend himself. By using the
previous definition, we can define the term of cyberbullying such as an
aggressive or intentional act or behavior that is carried out in a repeatedly
by using an electronic form of contact. It can be said that cyberbullying is
the evolution of bullying.


 The Internet has become the new Wild West of the 21st century and has to be explored because it provides excitement and adventure, thus as Wild West, Internet is full of dangers. Also, \cite{Franek} 
mentioned that \textquotedblleft We need to be sheriffs in this new Wild West
-- a cyber -- world buzzing with kids just a few keystrokes away from harming
other people, often for no other reason than that the sheriffs are sleeping.
As anyone who has ever been a victim of bullying and harassment will tell you
the bullets may not be real, but they can hurt\textquotedblright%
\ .

\section{Bullying}

According to \cite{Olweus1993a}, 
bullying can be defined as a student
is being bullied or victimized when he or she is exposed, repeatedly and over
time, to negative actions on one or more students. A negative action, with the
aggressive behavior definition, when someone intentionally inflicts or attempt
to inflict, injury upon another \cite{Olweus1973b}. In bullying, there is the
physical contact with bully and victim, in which negative actions carried out
by words or in other ways, for examples such as making grimaces or unpleasant
gestures, spreading rumors and intentional excluse him/her from a group. 

\subsection{Bullies aggressive reaction}

According to \cite{Olweus1996}, there are four
factors that affect bullies aggressive reaction. These factors are the answers
to the question \textquotedblleft What kind of rearing and other conditions
during childhood are conductive to the development of an aggressive reaction
pattern?\textquotedblright.

The factors that mentioned before are:

\begin{enumerate}
 \item The basic emotional attitude of the caretaker, who usually is the
mother, to the child during the early years. A negative emotional attitude can be characterized by the absence of warmth and involvement. 

 \item The aggressive behavior level of the child is increasing when there
is a tolerant attitude from the caretaker. Caretakers have to set clear limits when there is an aggressive behavior towards peers, brothers and sisters, and other adults. Power-assertive methods which are using to child upbringings such as physical punishment and violent emotional outbursts are likely to make children more aggressive than the average child. In other words, \textquotedblleft violence begets violence\textquotedblright.
 \item Physical punishment and violent emotional outbreaks methods are
likely to affect children to have an aggressive attitude. When parents use
those kinds of methods, children are more likely to become aggressive. It is
already known that "violence begets violence"
 \item The way of treatment to a child is inherited. A child who has a
\textquotedblleft short-tempered\textquotedblright\ temperament is more likely to be an aggressive youngest than a child with a common temperament. 

\end{enumerate}

According to above factors, it can be deduced that the child who had too
little love and care, and too much freedom in his childhood is more likely to
become aggressive. Thus, the socio-economic conditions of the family are not
related to the aggressive behavior of the child.

\section{Cyberbullying}

 Cyberbullying is the evolution of traditional bullying by using the
internet technology. According to researchers, cyberbullying refers to
bullying of others by using mobile phones and the internet \cite{Smith}. Such as bullying, can take part in all countries,
without being relevant to the culture and the religion of the victim \cite{Berger}.


The anonymity, the invasion of personal life, and the fact that the
victim cannot hide from torturers in the electronic world are characteristics
that make electronic bullying more painful to the victims \cite{Mishna}.

\subsection{Cyberbulling and parenting}

The role of parenting in cyberbullying is something new which cannot
provide us a lot of information.

Children exclude their parents from internet activities because it is
considered that their privacy is valued \cite{Subrahmanyam}.
Thus, cyberbullying is not as visible as traditional bullying. Furthermore,
young people are not telling their parents about the involvement in
cyberbullying because they are afraid of the punishment, the loss of computer
privileges and the isolation of peers (\cite{Bath}, \cite{Kraft}, \cite{Mishna}). These means that parents are often unaware of their child
is a cyberbully or is cyberbullied (\cite{Aricak}, \cite{Dehue}). 

There are studies that examined the relationship between parenting
characteristics and cyberbullying. They found that children who cyberbully
experience limited parental monitoring, stronger parental discipline and a
weaker emotional bond with their parents than children who do not cyberbully
\cite{Ybarra},
\cite{Wang}.

Insufficient parenting reduces the social competence of their children
as an example, the ability to develop positive friendships. On the contrary,
parents who oversee their children and get involved in them, the aggressive
behavior of the children within and without family is more likely to be
reduced \cite{Duman}, \cite{Knutson}, \cite{Mazefsky}. The way that parents interact with their children
affects children to interact with others in the same way. If parents interact
with their children in a hostile and cold manner, encourage them to use the
same manner in their interactions. This affect children socialization and
children are more likely to be a bully \cite{Pontzer}.

Anti-social behavior is unlikely to happen in a child who has grown up
with supervision, discipline, and affection (\cite{Bacchini}, \cite{Brown}, \cite{Demetriou}, \cite{Getachew}, \cite{Knutson}, \cite{Luyckx}).

There were a lot of researchers who investigated parental
responsiveness and traditional bullying. A negative association is the result
of these elements relation in which as parents were less responsive, their
children are more likely to have a bullying behavior and victimization is
increased (\cite{Flouri}, \cite{Georgiou}, \cite{Ok}). Some
other researchers compared bullies, bully-victims and victims and found that
bullies and bully-victims experienced less parental responsiveness than
victims (\cite{Demaray}, \cite{Smith2}, \cite{Stevens}).

\begin{figure}[h]
\centering
\includegraphics[width=0.7\linewidth]{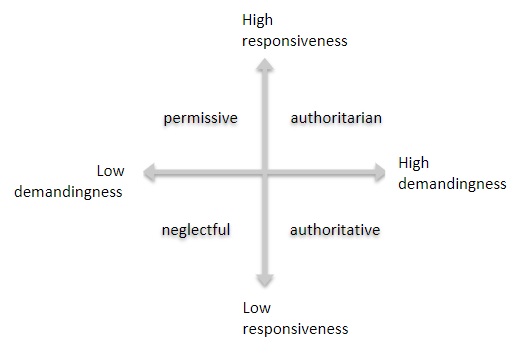} 
\caption{Parenting styles}
\label{fig:f1}
\end{figure}

Both dimensions however co-exist and influence each other \cite{Baumrind}, \cite{Spera}. Figure \ref{fig:f1}, shows the four parenting styles and the
combination of both dimensions results \cite{Dehue}. 

Various researches examined the relation between the characteristics of
parenting and cyberbullying. The results were that children with cyberbully
experience had limited parental monitoring, stronger parental discipline and
weaker emotional bond with their parents, in contrast to children with no
cyberbully experience \cite{Ybarra}, \cite{Wang},\cite{Wong}.

\subsection{Cyberbullying and Traditional Bullying}

Traditional bullying and cyberbullying have the same features which
are repetitiveness, intentionality and power balance. In the case of
cyberbullying, is appeared the risk of misunderstanding as is bereft of
non-verbal cues. This means that a joke may be misunderstanding and could be
perceived as intentional and hurtful. Further, the cyberbullying perpetrators
may be unknown, contrary to traditional bullying \cite{Dehue}, \cite{Dempsey}, \cite{Huang}, \cite{Ybarra}, \cite{Dehue2}.

\section{Survey Methodology}

This section presents the design of the survey, the process of
designing the questionnaire, its structure, the sampling method, the sample as
well as the data analysis techniques used.

 Non-probable sampling method was applied. The data was collected via an
electronic questionnaire which was developed based on extended literature
review. Specifically, it was developed by the use of Google forms and
distributed via e-mail and social media. Our questions based on the sub-scales
from the bully/Victim Questionnaire, and parenting with an adjusted version of
the parenting style Questionnaire.

Regarding the respondents, anyone could complete it anonymously with
our aim to ensure confidentiality.

In the present survey, the primary data collection method, which is
known as a quantitative method and based on a sample survey using a
standardized questionnaire, was also selected.

This questionnaire included closed-ended questions. Closed type
questions are the questions that are accompanied by a series of suggested
answers to the respondent from which to select one. In the present
questionnaire, there are "yes-no" questions, multiple choice questions, and
Likert scale questions. In the Likert scale questions the Likert scale with 5
grades was used, where option 1 does not indicate any option and option 5 means very high.

\subsection{Questionnaire Structure}

The questionnaire consists of six units. The first unit consists of
the demographic questions. 

In the second, questions about internet use and the reasons for were
applied. The third unit consists of questions about cyberbullying.  In this
section, the form, the feelings, the consequences and the way that came up
against the cyberbullying phenomenon were asked to be completed. 

The next unit consists of questions about traditional bullying. In
this section, the respondents asked to answer questions about the genre of
traditional bullying, if victims had a solution to this problem and in which
way. 

The fifth unit consists of questions about cyber bullies. In this
unit, respondents have to answer if they became bullies, for which the reasons and in each way. 

Finally, in  the sixth section there was a validation question in
order to verify that a human completed the measurement item and not a
machine.

The questions about cyberbullying and traditional bullying were
adjusted and based on \cite{Olweus1993a} questionnaire, and questions about the
parenting dimensions were measured by using adjusted questions of the
parenting style questionnaire of \cite{Exter}.

Prior to the study, the questionnaire was pre-tested by 20 students who
did not participate in the study. As a result of the pretest, some questions
were adjusted to facilitate clarity, understanding,and ease in answering the  questions.

\subsubsection{Research Questions}

\begin{itemize}
 \item To what extent is cyberbullying represented in Greece?
 \item Which gender is more vulnerable to bullying?
 \item Which is the family relationship associated with the cyberbullying
victims?
 \item Is cyberbullying associated with the hours spent on the internet?
 \item Is there an association between traditional bullying victims and
cyberbullying victims?
\item Is there an association between cyberbullying victims and
bullies?
\end{itemize}

Data gathering period was from March up to April 2018. 525 responses
were collected; but only 466 of them were valid. Afterwards, data were
encoded and analysed using the SPSS Statistical program.

\section{Main Results}

\subsection{Gender Results}

The sample consisted of 466 valid questionnaires.

Of 466 participants, 282 participants (60.5\%) were women and 184
(39.5\%) were men.

\subsection{Age results }

About the age of the participants we had: 

\begin{itemize}
 \item 128 (27.5\%) participants  were under the age of 18
 \item 191 (41\%) participants  were between 19-25
 \item 32 (6.9\%) participants  were between 26-30
 \item 38 (8.2\%)  participants  were between 36-40 
 \item 53 (11.4\%) participants were over the age of 40 years old
\end{itemize}

\subsection{Hours that spends online results}

About the hours that our respondents spend online, we have:

\begin{itemize}
 \item  193 (41.4\%) of our participants  spent 1-3 hours
online 
 \item 179 (38.4\%) of our participants spend 4-6 hours
online
 \item 64 (13.7\%) of our participants spend 7-9 hours
online
 \item 30 (6.4\%) of our participants spend more than 9 hours
online
\end{itemize}

\subsection{Gender and cyberbullying victims results}

In order to examine which gender is more vulnerable to cyberbullying,
we used cross-tabulation in SPSS. The results are:

\begin{itemize}
 \item 61 of our respondents have been cyberbullying
victims.
 \item 18 cyberbullying victims were male (29.5\%).
 \item 42 cyberbullying victims were female (70.5\%).
\end{itemize}

One of the research questions was to identify which gender is more
vulnerable to cyberbullying. The answer to this question is that women are
more vulnerable to cyberbullying as 70.5\% of cyberbullying victims were
women, in contrast to 29.5\% of men victims.

\subsection{Age and cyberbullying victims}

Getting results about the age of cyberbullying victims, SPSS
cross-tabulation was used. 

The results were:

\begin{itemize}
 \item \textless 18 age scale, 14 of our respondents were cyberbullying victims (23\%)
 \item 19-25 age scale, 31 of our respondents were cyberbullying
victims (50.8\%).
 \item 26-30 age scale, 4 of our respondents were cyberbullying
victims (6.6\%).
 \item 31-35 age scale, 4 of our respondents were cyberbullying
victims (6.6\%).
 \item 36-40 age scale, 2 of our respondents were cyberbullying
victims (3.3\%).
 \item \textgreater 40 age scale, 6 of our respondents were cyberbullying victims (9.7\%).
 
\end{itemize}

Our survey showed that the weakest age group is the 19-25 with 31
participants to be victims of cyberbullying. Continuing, the age group with
the less cyberbullying victims is the age group of 36-40 with 0.4\%.

\subsection{Cyberbullying victims and hours that spend online results}

SPSS cross-tabulation was used in order to examine the hours that
cyberbullying victims spend online. Our survey showed that most of the
cyberbullying victims (41\%) spend 4-6 hours online. The next following
percentage is 27.9\% where cyberbullying victims spend 1-3 hours online. So the answer to our research question is that most cyberbullying victims spend online 4-6 hours.

\subsection{Cyberbullying victims and faced the problem results.}

In order to examine if cyberbullying victims faced their problem, we used
cross-tabulation in SPSS. The results that we have from crosstabulation were:

\begin{itemize}
 \item 48 of the cyberbullying victims (78.7\%) faced up
their problem.
 \item 13 of the cyberbullying victims (21.3\%) didn't find
a solution to their problem.
\end{itemize}

\subsection{Cyberbullying victims and traditional bullying victims results}

In order to examine the number of cyberbullying victims and
traditional bullying victims, cross-tabulation was used in SPSS. The results showed us: 

\begin{itemize} 
 \item 291 (62.4\%) respondents haven't been cyberbullying victims
and haven't been traditional bullying victims as well.
 \item 114 (24.5\%)  respondents haven't been cyberbullying victims
but have been traditional bullying victims.
 \item 28 (6\%)  respondents have been cyberrbullying victims but
they haven't been traditional bullying victims.
 \item 33 (7.1\%) respondents have been cyberbullying victims and
traditional bullying victims too.

\end{itemize}

\subsection{Cyberbullying victims and traditional bullying victims}

\begin{figure}[h]
\centering
\includegraphics[width=0.7\linewidth]{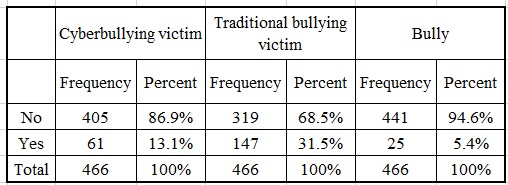}
\caption{Cyberbullying victims,traditional victims,and bullies}
\label{fig:t1}%
\end{figure}

Figure \ref{fig:t1} shows that 13.1\% of our respondents were
cyberbullying victims in contrast to the traditional bullying victims who were the 31.5\%. Also, 5.4\% of our participants bullied other people. 

According to Figure \ref{fig:t1}, the answer to our research inquiry about the
extension of cyberbullying in Greece is that, cyberbullying appears in half of
the proportion compared to the traditional bullying. 

\subsection{Cyberbullying victims and bullies}

According to Figure \ref{fig:t1}, 61 of our respondents have been cyberbullying victims and 8 respondents have been bullying someone else too.

\begin{figure}[h]
\centering
\includegraphics[width=0.7\linewidth]{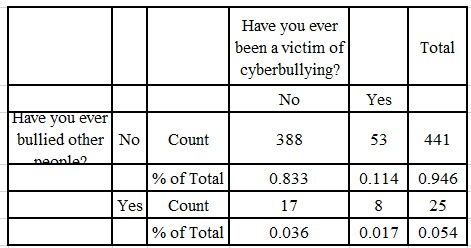} 
\caption{Cyberbullying victims/bullies}
\label{fig:t2-1}
\end{figure}

\begin{figure}[h]
\centering
\includegraphics[width=0.7\linewidth]{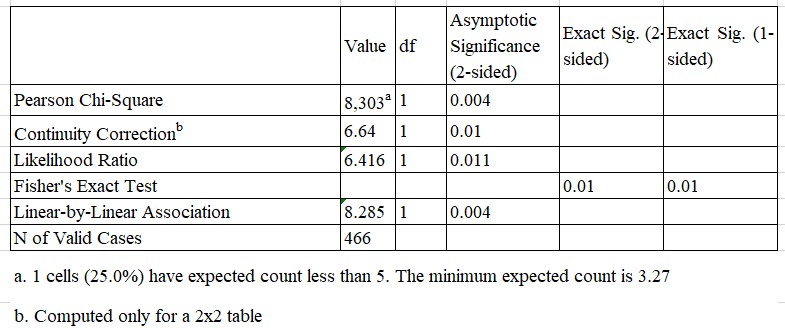} 
\caption{Chi-Square Tests}
\label{fig:t2-2}
\end{figure}

\begin{figure}[h]
\centering
\includegraphics[width=0.7\linewidth]{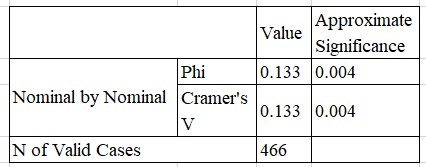}
\caption{Symmetric Measures}
\label{fig:t2-3}
\end{figure}

From the Figures \ref{fig:t2-1}, \ref{fig:t2-2} and  \ref{fig:t2-3}, we can see that the significance value is lower 
than the \textquotedblleft a\textquotedblright\ value (0.004%
$<$%
0.05) and there is a statistical dependence between the variables on the level of significance a=0.05. Statistically significant dependence was identified and it is following the assessment of its intensity.
For this reason, we used the causation coefficients Phi Cramer's V  in order to compare the strength of the coefficient between the variables. Phi and Cramer's V are estimated factors of the intensity (size) of the connection between two quality variables. In the case of independence between the two variables, the value of the factors is close to zero (0). In our case, Phi Coefficient is 0.133 and this means that there is a relationship between bullies and cyberbullying victims.

Finally, in order to determine the existence of dependence between bullied
other people and cyberbullying victims, the Chi-square test of independence was used to determine if there is a significant relationship between the two variables. Also, Phi and Cramer's V coefficients were calculated. According to the results, there is a significant relationship on the Level of significance a=0.05 between the bullied other people and cyberbullying victims(x2=8.303,  sig=0.004). As Phi dependence coefficient is 0.133, there is a small dependence between the bullied other people and the cyberbullying victims.

\subsection{Cyberbullying victims and traditional bullying victims}

\begin{figure}[h]
\centering
\includegraphics[width=0.7\linewidth]{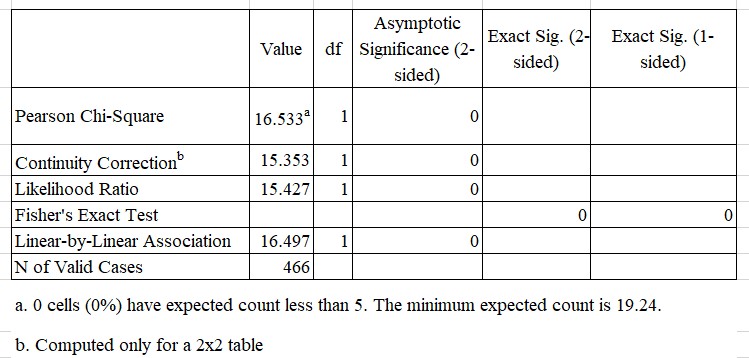}
\caption{Traditional and cyber bullying Chi-Square Tests}
\label{fig:t3-2}
\end{figure}

\begin{figure}[h]
\centering
\includegraphics[width=0.7\linewidth]{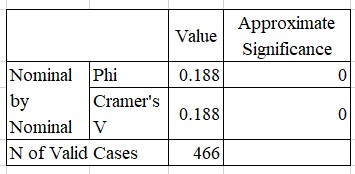} 
\caption{Symmetric Measures}
\label{fig:t3-3}
\end{figure}

According to Figures \ref{fig:t3-1}, \ref{fig:t3-2} and \ref{fig:t3-3}  respondents have been traditional bullying victims and 33 respondents have been cyberbullying victims too.

We can see that the sig. value is lower than the
\textquotedblleft a\textquotedblright\ value (0.000%
$<$%
0.05). According to this, the null hypothesis is rejected and there is a
statistical dependence between the variables on the level of significance a=0.05:

\begin{figure}[h]
\centering
\includegraphics[width=0.7\linewidth]{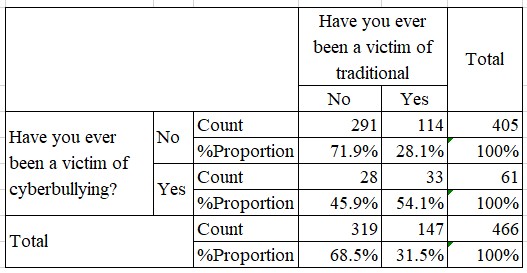}
\caption{Traditional and cyber bullying Cross-tabulation}
\label{fig:t3-1}
\end{figure}


Statistically, significant dependence was identified and it is
following the assessment of its intensity.

For this reason, we are using the causation coefficients Phi Cramer's V
in order to compare the strength of the coefficient between the variables.

Phi Coefficient is 0.188 and this means that there is a relationship
between traditional bullying victims and in cyberbullying victims.

Finally, for determining the existence of dependence between
traditional bullying victims and cyberbullying victims, the Chi-square test of independence was used to determine if there is a significant relationship between the two variables. Also, Phi and Cramer's V coefficients were calculated. According to the results, there is a significant relationship in the Level of significance a=0.05 between the bullied other people and cyberbullying victims(x2=16.533, sig.=0.004). As Phi dependence coefficient is 0.188, there is a small dependence between the bullied other people and the cyberbullying victims. Therefore, we reject the null hypothesis.



\begin{figure}[h]
\centering
\includegraphics[width=0.7\linewidth]{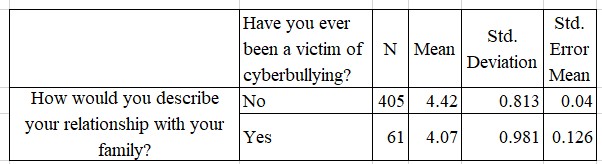} 
\caption{Group Statistics}
\label{fig:t4-1}
\end{figure}

\begin{figure}[b]
\centering
\includegraphics[width=0.7\linewidth]{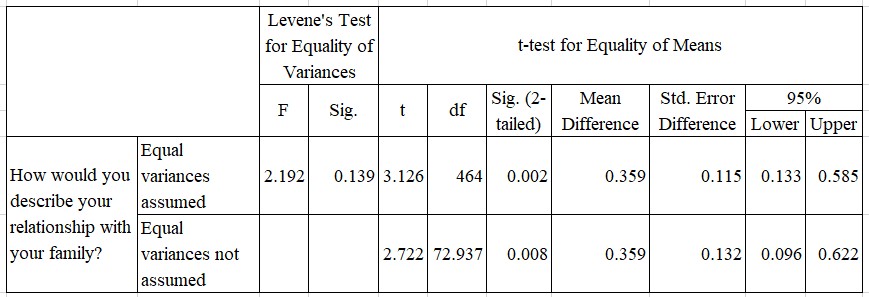} 
\caption{Independent Samples Test}
\label{fig:t4-2}
♠\end{figure}

According to Figures \ref{fig:t4-1} and \ref{fig:t4-2}, we have the following research hypothesis.

H$_{ 0}$: There is no difference between the family
relation with the cyberbullying victims

 H$_{1}$: There is a difference between the family
relation with the cyberbullying victims

Also, we have the following research hypothesis.

 H$_{00}$: Family relation is not related to the
cyberbullying victims

 H$_{10}$: Family relation is related to the
cyberbullying victims

Sig value of the Levene's Test for Equality of Variances is 0.139(%
$>$%
0.05). This means that we reject hypothesis H$_{10}$ and we
accept H$_{10}$ hypothesis. 

Also, we accept the hypothesis H$_{1} $  because of
Sig. (2-tailed) value is 0.002 (%
$<$%
0.05). This means that there is a correlation between the cyberbullying
victims and the family relation. From Figure \ref{fig:t4-1}  we can see that respondents
who have been cyberbullying victims were less well connected with their family
(M=4.07) compared with the respondents who were cyberbullying victims
(M=4.42). Therefore, we can say that family relation is related to
cyberbullying victimization.

\section{Conclusions}

Cyberbullying is based on the bullying by using the technology. Also, technology offer opportunities and can be painful for those who are targets of cyberbullying.

Our survey aimed to explore the extent of bullying and cyberbullying in Greece. In order to answer the research questions, we collected data which analyzed by SPSS statistical program. The important results that we have are:

\begin{itemize}
 \item Cyberbullying has a small extent in Greece, in
contrast to traditional bullying which has a larger area.
 \item Girls are more vulnerable to cyberbullying.
 \item Cyberbullying victims spend 4-6 hours online.
 \item There is a statistical dependence between
cyberbullying victims and traditional bullying victims.
 \item There is a correlation between cyberbullying victims
and family relation.

\end{itemize}

From our research results, it can be said that parents and schools need to help and prepare children from the hazards of the new technology as cyberbullying will evolve as technology continues to evolve. Future studies need to explore a larger sample size. Also elements such as cyberbullying and bullying relation and the parenting styles of cyberbullying and bullying victims need to be explored deeper.

\end{document}